\documentclass[aps,prd,twocolumn,preprintnumbers,letterpaper,superscriptaddress,nofootinbib,floatfix]{revtex4-2}
\pdfoutput=1

\usepackage{graphicx}
\usepackage{microtype}
\usepackage{amsmath}
\usepackage{amssymb}
\usepackage{subfigure}
\usepackage{hyperref}
\usepackage{url}
\usepackage{xcolor}
\usepackage{color}
\usepackage{mathrsfs}
\usepackage{amsfonts}
\usepackage{latexsym}
\usepackage{ragged2e}
\usepackage{epstopdf}
\usepackage{textcomp}
\usepackage{phaistos}
\usepackage[utf8]{inputenc}
\usepackage[normalem]{ulem}
\usepackage{float}

\graphicspath{{Images/}}
\hypersetup{
    colorlinks=true,
    linkcolor=blue,
    citecolor=magenta,
    urlcolor=red
}

\newcommand{\dd}{\mathop{}\!\mathrm{d}}
\newcommand{\Boxop}{\Box}
\newcommand{\Mop}{\mathscr M}
\newcommand{\Pop}{\mathscr P}
\newcommand{\PT}{\Pi^{\mathrm T}}
\newcommand{\ret}{\mathrm{ret}}
\newcommand{\ii}{\mathrm i}
\newcommand{\Lie}{\mathcal L}

\begin{document}

\title{
Retarded Green Function for Bopp–Podolsky Electrodynamics in Four-Dimensional de Sitter Spacetime
}

\author{Wei \surname{Fan}}
\email{fanwei@just.edu.cn}
\affiliation{School of Science, Jiangsu University of Science and Technology, 212100, Zhenjiang, China.}

\author{Xiao Yan \surname{Chew}}
\email{xiao.yan.chew@just.edu.cn}
\affiliation{School of Science, Jiangsu University of Science and Technology, 212100, Zhenjiang, China.}

\begin{abstract}
{ 
We study classical Bopp--Podolsky (BP) electrodynamics on
four-dimensional de Sitter spacetime, focusing on its transverse
retarded Green function and gauge-invariant electromagnetic response.
Because de Sitter spacetime is maximally symmetric, its vector Green
functions admit a covariant expansion in a finite basis of invariant
bitensors, with scalar coefficients depending only on the invariant
separation. For conserved sources, the fourth-order field operator
factorizes into Maxwell and massive-vector operators, yielding the BP
retarded kernel as the difference of their transverse Green functions.
Using the invariant bitensor decomposition, we reduce the propagator
equations to a hypergeometric master equation and determine the
distributional structure of its solution. The common leading
short-distance singularity of the Maxwell and massive-vector sectors
cancels, leaving a logarithmic master singularity and a softened
vector-potential wavefront. After one curl, the gauge-invariant
field-strength kernel contains a null-cone delta distribution and a
regular timelike tail. The smooth Maxwell master tail is annihilated
by the curl, so the interior electromagnetic tail is generated
entirely by the massive sector. We also derive an exact
horizon-regular static potential, demonstrate the cancellation of the
Coulomb divergence, recover the flat-space BP potential, and formulate
the generalized Gauss law. On global de Sitter spacetime, the
compactness of the spatial sections requires the total charge to
vanish; we implement this constraint using an antipodal charge
configuration with smooth factorized initial data. The analysis is
classical and does not assume a quantum Lee--Wick contour or a
preferred de Sitter vacuum.} 
\end{abstract}

\maketitle

\section{Introduction}

%Higher-derivative field theories provide a useful setting in which ultraviolet behavior can be modified while covariance and gauge symmetry remain explicit.  One of the earliest examples is Bopp--Podolsky electrodynamics, introduced independently by Bopp \cite{Bopp1940} and Podolsky \cite{Podolsky1942} and subsequently developed as a linear generalization of Maxwell theory \cite{PodolskySchwed1948}.  Its Lagrangian supplements the Maxwell term by a gauge-invariant contribution quadratic in the divergence of the field strength.  The equations are fourth order in the vector potential but remain linear and locally gauge invariant.

Higher-derivative electrodynamics was originally motivated by the divergent potential, field energy, and self-interaction associated with an ideal point charge in Maxwell theory. One of the earliest such modifications is Bopp--Podolsky (BP) electrodynamics, introduced independently by Bopp \cite{Bopp1940} and Podolsky \cite{Podolsky1942} and subsequently developed as a linear generalization of Maxwell theory \cite{PodolskySchwed1948}. Its Lagrangian supplements the Maxwell term with a gauge-invariant contribution quadratic in the divergence of the field strength, thereby introducing a characteristic inverse-length scale \(M\). Although the resulting equations are fourth order in the vector potential, the theory remains linear, local, Lorentz covariant, and gauge invariant. In flat-space electrostatics, the cancellation between its massless and massive sectors removes the Coulomb divergence of the point-charge potential and renders the electrostatic field energy finite. This classical regularization provides a principal motivation for examining whether the same cancellation mechanism persists on curved backgrounds. It does not, however, imply complete smoothness for arbitrary dynamical sources, for which weaker directional or distributional singularities may remain.

The mechanism underlying this regularization becomes transparent through the factorization of the flat-spacetime fourth-order operator into a massless Maxwell sector and a massive-vector sector. Accordingly, the BP propagator is given by the difference between the corresponding massless and massive second-order propagators. Their leading short-distance singularities cancel, changing the large-momentum behavior from \(p^{-2}\) to \(p^{-4}\) and softening the point-charge field. The resulting causal Green functions, electromagnetic fields, and light-cone structure have been studied systematically in flat spacetime \cite{Lazar2019}. The implications of this regularization for the classical self-force have also been reviewed in comparison with nonlinear and other higher-derivative theories \cite{Perlick2015}, while a detailed distributional treatment established the conditions under which point-charge fields and the associated self-force remain finite \cite{Gratus2015}. Additional phenomenological consequences include constraints on the characteristic Bopp length from the hydrogen spectrum \cite{CarleyKiesslingPerlick2019}. More recently, the transverse and longitudinal wave sectors of Bopp--Land'e--Thomas--Podolsky electrodynamics have been analyzed in both vacuum and plasma environments \cite{ShalaPerlick2025}.

Extending this factorized construction to curved spacetime is not straightforward. Covariant derivatives do not commute, and flat-space expressions that are equivalent up to integrations by parts can lead, after covariantization, to theories differing by explicit curvature couplings. Curved-space formulations of BP electrodynamics, including their Maxwell--massive decomposition in static black-hole geometries, have been investigated previously \cite{Cuzinatto2018}. In the present work, we adopt the minimal covariantization obtained by replacing ordinary derivatives with covariant derivatives without introducing additional nonminimal curvature couplings. On an Einstein background, and after restriction to the transverse sector selected by a conserved current, the resulting fourth-order vector operator factorizes into massless Maxwell and massive-vector operators. Therefore, de Sitter spacetime preserves the essential factorization mechanism responsible for the flat-space regularization, while introducing nontrivial effects associated with curvature, causal tails, and global topology.

De Sitter spacetime is a maximally symmetric Lorentzian manifold with constant positive curvature. This maximal symmetry ensures that a de Sitter-invariant vector two-point function can be expanded in a finite basis of invariant bitensors constructed from the metric, the geodesic separation, and the parallel propagator. Its tensor coefficient functions consequently depend only on a single invariant variable, such as \(Z(x,x')\), reducing the propagator equations to ordinary differential equations. Beyond this technical advantage, de Sitter spacetime is physically motivated as the idealized limit of two important cosmological regimes: the approximately de Sitter expansion expected during primordial inflation and the asymptotic de Sitter behavior of a late universe dominated by a positive cosmological constant. Therefore, it provides a useful setting for studying both primordial quantum fields and the long-distance propagation of fields in an accelerating universe. Vector two-point functions on maximally symmetric spacetimes have been formulated in an invariant bitensor basis \cite{AllenJacobson1986}. A normalized transverse massive-vector propagator with the appropriate flat-space and massless limits, together with a verified classical response, was subsequently obtained \cite{TsamisWoodard2007}. Mode-sum Stueckelberg constructions clarified the relation between the Proca and transverse formulations \cite{FroebHiguchi2014}, while photon propagators in covariant de Sitter gauges have also been investigated \cite{DomazetProkopec2014}.

\begin{table*}[t]
\caption{Representative de Sitter propagator constructions relevant to the invariant-bitensor program.  ``Established'' refers to the free two-point problem; interacting entries are perturbative and should not be read as exact dressed propagators.}
\label{tab:dspropagators}
\begin{ruledtabular}
\begin{tabular}{p{0.19\textwidth}p{0.28\textwidth}p{0.46\textwidth}}
Theory & Status on de Sitter & Representative result\\
\hline
Klein--Gordon scalar & Extensive & Invariant Wightman/Feynman functions and causal Green functions \cite{ChernikovTagirov1968,BunchDavies1978,BrosGazeauMoschella1994,BrosMoschella1996}\\
Dirac field & Extensive & Spinor two-point function in a spin-parallel-propagator basis \cite{AllenLutken1986,BartesaghiEtAl2001}\\
Maxwell theory & Established; gauge/state subtleties & Covariant-gauge and transverse photon kernels \cite{AllenJacobson1986,DomazetProkopec2014,GlavanProkopec2023}\\
Proca theory & Established & Massive transverse-vector propagator \cite{AllenJacobson1986,TsamisWoodard2007}\\
Stueckelberg electrodynamics & Established & Massive-vector propagator with gauge parameter \cite{FroebHiguchi2014}\\
Linearized gravity & Extensive but gauge-sensitive & Rank-$(2,2)$ graviton two-point functions in several gauges \cite{AllenTuryn1987,HiguchiKouris2001,FroebHiguchiLima2016}\\
Yang--Mills theory & Free kernel established; interacting theory perturbative & Maxwell-like color-diagonal kernel plus ghost sector, and global-de Sitter three-gluon amplitudes from transverse $S^3$ harmonics \cite{ShaikhFaizalShah2023,TaylorZhu2026Gluon}\\
Scalar QED & One-loop vacuum polarization & Effective photon equations and partial dressing \cite{ProkopecTornkvistWoodard2002,LeonardProkopecWoodard2013}\\
QED with graviton loops & One-loop vacuum polarization & Perturbative correction to photon mode functions \cite{GlavanMiaoProkopecWoodard2017}\\
\end{tabular}
\end{ruledtabular}
\end{table*}

The invariant bitensor formalism is not restricted to vector fields. For a scalar field, the propagator is a biscalar and, on a maximally symmetric background in a de Sitter-invariant state, depends only on the invariant separation \(Z(x,x')\), or an equivalent variable such as \(y(x,x')\). Early constructions established the invariant scalar two-point functions and the state now known as the Euclidean or Bunch--Davies state \cite{ChernikovTagirov1968,BunchDavies1978}. De Sitter quantum field theory was subsequently formulated in terms of holomorphic functions on the complexified de Sitter manifold and coordinate-independent de Sitter plane waves \cite{BrosGazeauMoschella1994}. Within this analytic framework, Wightman and Feynman two-point functions arise as boundary values in appropriate complex domains, while the usual Minkowski spectral condition is replaced by a de Sitter analyticity condition \cite{BrosMoschella1996}. Recent discussions further emphasize that this invariant construction does not require a preferred coordinate patch or a special coordinate separation \cite{Moschella2024}. The Feynman, Wightman, retarded, and advanced kernels are then related through their boundary values and discontinuities. Analogous invariant constructions for spinor fields express the propagator in terms of scalar coefficient functions multiplying the spin parallel propagator and its contraction with the geodesic tangent \cite{AllenLutken1986,BartesaghiEtAl2001}. For spin-one fields, the parallel propagator and the product of the two geodesic tangent vectors form the standard two-element bitensor basis underlying the Proca, Stueckelberg, photon, and transverse massive-vector propagators discussed above. Nevertheless, outside an exactly transverse formulation, the photon two-point function can exhibit subtleties associated with gauge fixing and possible de Sitter-symmetry breaking \cite{GlavanProkopec2023}.

Complementary recent work formulates scattering directly on global de Sitter spacetime in terms of $SO(1,4)$ representations, compact $S^3$ spatial slices, and a controlled flat-space limit \cite{TaylorZhu2025}.  Its scalar continuation highlights genuinely de Sitter deep-infrared effects \cite{TaylorZhu2026Scalar}.  For gauge fields, transverse coexact covector harmonics on $S^3$ have been used to construct tree-level three-gluon amplitudes \cite{TaylorZhu2026Gluon}.  These studies concern asymptotic states and amplitudes rather than the causal inverse constructed below, but they reinforce the importance of global topology and the transverse mode content in de Sitter gauge theory.

Table~\ref{tab:dspropagators} places the present calculation in the wider de Sitter propagator literature \cite{AllenJacobson1986,TsamisWoodard2007,FroebHiguchi2014,ChernikovTagirov1968,TaylorZhu2026Gluon,BunchDavies1978,BrosGazeauMoschella1994,BrosMoschella1996,DomazetProkopec2014,AllenLutken1986,AllenTuryn1987,HiguchiKouris2001,FroebHiguchiLima2016,BartesaghiEtAl2001,ProkopecTornkvistWoodard2002,LeonardProkopecWoodard2013,GlavanMiaoProkopecWoodard2017,ShaikhFaizalShah2023}.  The last two rows require care: they concern loop-corrected photon propagation through a vacuum-polarization kernel, not a new exact free propagator.  In scalar QED the renormalized one-loop polarization of a light or massless charged scalar has been computed and used to obtain effective photon equations \cite{ProkopecTornkvistWoodard2002,LeonardProkopecWoodard2013}.  Graviton contributions have likewise been used to calculate perturbative corrections to dynamical photon mode functions \cite{GlavanMiaoProkopecWoodard2017}.  For Yang--Mills theory the color-diagonal free gauge-field kernel is inherited from the Maxwell sector, whereas covariant perturbation theory also requires ghost propagators and an infrared prescription \cite{ShaikhFaizalShah2023}; this is less complete than the free scalar, spinor, or Proca cases.

For spin two, the propagator is expanded in rank-$(2,2)$ maximally symmetric bitensors built from the metric, parallel propagator, and geodesic tangents.  Closed de Sitter graviton propagators obtained by this method include the Euclidean-vacuum construction of Allen and Turyn \cite{AllenTuryn1987}, covariant-gauge representations \cite{HiguchiKouris2001}, and later mode-sum treatments in the Poincar\'e patch \cite{FroebHiguchiLima2016}.  Antisymmetric tensor and differential-form propagators can be treated similarly, although parity-odd or self-dual form theories may require additional invariant structures beyond the usual metric-and-tangent basis \cite{BenaNastaseVaman2000}.  The present work belongs to the spin-one branch of this program: its new ingredient is the higher-derivative Maxwell--massive subtraction and the associated retarded distributional analysis, rather than a new classification of maximally symmetric bitensors.

These works supply the required second-order ingredients, but their BP combination has a distinct distributional structure.  The transverse source projector, relative normalization, causal prescription, and gauge-invariant curls must be tracked consistently.  Furthermore, compact global de Sitter slices impose a vanishing-total-charge constraint, so a static-patch point charge requires a globally neutral completion.

We construct the classical transverse BP Green function on four-dimensional de Sitter spacetime.  We derive its invariant master function, retarded discontinuity, and softened wavefront; obtain one- and two-curl gauge-invariant kernels; and provide flat, static, and global consistency checks.  The analysis is deliberately classical.  The negative-residue massive pole is noted, but quantum unitarity, Lee--Wick contour choices, and vacuum selection are outside our scope.

%\section{Action, equations, and factorization}

\section{Bopp-Podolsky Theory}

We begin with the minimally covariantized Bopp--Podolsky (BP) action and
adopt the metric signature $(-+++)$:
\begin{equation}
 S=\int\!\dd^4x\sqrt{-g}\left[
 -\frac14F_{\mu\nu}F^{\mu\nu}
 -\frac{1}{2M^2}B_\mu B^\mu
 -J_\mu A^\mu
 \right].
 \label{eq:action}
\end{equation}
Here $A_\mu$ is the gauge field,
$J^\mu$ is a conserved current,
\begin{equation}
 \nabla_\mu J^\mu=0,
 \label{eq:currentconservation}
\end{equation}
and $M>0$ is the BP mass scale. Equivalently,
$M^{-1}$ defines the characteristic BP length in units
with $\hbar=c=1$. The electromagnetic field strength and its
covariant divergence are
\begin{equation}
 F_{\mu\nu}=2\nabla_{[\mu}A_{\nu]},
 \qquad
 B^\nu=\nabla_\mu F^{\mu\nu}.
 \label{eq:definitions}
\end{equation}
Thus, $B^\mu$ is a derived vector constructed from $A_\mu$, rather
than an independent field. Variation of the action with respect to
$A_\nu$ gives
\begin{equation}
 B^\nu-\frac1{M^2}\nabla_\mu
 \left(\nabla^\mu B^\nu-\nabla^\nu B^\mu\right)
 =J^\nu.
 \label{eq:eom}
\end{equation}

To expose the differential operator acting on the vector potential, we
rewrite $B^\mu=\nabla_\nu F^{\nu\mu}$ using
Eq.~\eqref{eq:definitions}.  We denote the covariant d'Alembertian
acting on a vector by
\begin{equation}
 \Boxop A^\mu\equiv\nabla_\alpha\nabla^\alpha A^\mu.
\end{equation}
Expanding the field strength gives
\begin{align}
 B^\mu
 &=\nabla_\nu\left(\nabla^\nu A^\mu-\nabla^\mu A^\nu\right)
 \nonumber\\
 &=\Boxop A^\mu-\nabla^\mu(\nabla_\nu A^\nu)
   -R^\mu{}_\nu A^\nu,
 \label{eq:Bpotential}
\end{align}
where the Ricci-tensor term follows from the commutator of covariant
derivatives.  Therefore, it is convenient to define the transverse
Maxwell operator
\begin{equation}
 \Mop^\mu{}_\nu
 =\delta^\mu{}_\nu\Boxop-R^\mu{}_\nu.
 \label{eq:maxwelloperator}
\end{equation}
For a general vector potential,
\begin{equation}
 B^\mu
 =\Mop^\mu{}_\nu A^\nu-\nabla^\mu(\nabla_\nu A^\nu).
 \label{eq:BMaxwelloperator}
\end{equation}
For conserved sources, the physical response may be restricted to the
transverse sector.  Imposing the Lorenz condition
$\nabla_\mu A^\mu=0$ then gives
\begin{equation}
 B^\mu=\Mop^\mu{}_\nu A^\nu.
 \label{eq:Btransverse}
\end{equation}
In four-dimensional de Sitter spacetime, the curvature tensors satisfy
\begin{equation}
 R_{\mu\nu}=3H^2g_{\mu\nu},
 \qquad
 R=12H^2,
 \label{eq:dScurvature}
\end{equation}
where $H>0$ is the de Sitter curvature scale, related to the
cosmological constant by $\Lambda=3H^2$. Hence, the transverse Maxwell
operator defined in Eq.~\eqref{eq:maxwelloperator} reduces to
\begin{equation}
 \Mop^\mu{}_\nu
 =\delta^\mu{}_\nu\left(\Boxop-3H^2\right).
 \label{eq:dSMaxwelloperator}
\end{equation}
Here $\Boxop$ continues to denote the covariant d'Alembertian acting
on a vector field.

Because $B^\mu$ is the covariant divergence of an antisymmetric tensor,
it is identically transverse:
\begin{equation}
 \nabla_\mu B^\mu
 =\nabla_\mu\nabla_\nu F^{\nu\mu}=0.
 \label{eq:Btransverseidentity}
\end{equation}
Consequently,
\begin{align}
 \nabla_\mu\left(\nabla^\mu B^\nu-\nabla^\nu B^\mu\right)
 &=\Boxop B^\nu-R^\nu{}_\mu B^\mu
 \nonumber\\
 &=\Mop^\nu{}_\mu B^\mu.
 \label{eq:operatoronB}
\end{align}
In the Lorenz gauge, Eq.~\eqref{eq:Btransverse} gives
$B^\mu=\Mop^\mu{}_\nu A^\nu$. Substitution into
Eq.~\eqref{eq:eom} then yields
\begin{equation}
 \boxed{
 \Pop A
 \equiv
 \Mop\left(1-\frac{\Mop}{M^2}\right)A
 =J
 }.
 \label{eq:factoroperator}
\end{equation}
Equivalently,
\begin{equation}
 \Pop=-\frac1{M^2}\Mop(\Mop-M^2),
 \label{eq:factoredoperator}
\end{equation}
which displays explicitly the massless Maxwell factor $\Mop$ and the
massive-vector factor $\Mop-M^2$.

The factorization in Eq.~\eqref{eq:factoroperator} contains the two
second-order operators $\Mop$ and $\Mop-M^2$.  We define their transverse
Green functions $G_0  (x,x')=i\Delta_0 (x,x')$ and $G_M(x,x')=i\Delta_M(x,x')$ by
\begin{align}
 \Mop [i\Delta_0 (x,x') ]&=i\PT (x,x'),\label{eq:G0def}\\
 (\Mop-M^2)[i\Delta_M (x,x')]&=i \PT (x,x'),\label{eq:GMdef}
\end{align}
where $G_0(x,x')$ is the Maxwell Green function, $G_M(x,x')$ is the massive-vector
Green function, and $\PT(x,x')$ is the transverse identity kernel.  
Using the parallel propagator $g_{\mu\nu'}(x,x')$ and the scalar Green function $G_{sc}=i \Delta_{sc}(x,x')$,
\begin{align}
\Box_x (i\Delta_{sc}(x,x')) = i\delta_g(x,x'),
\end{align}
the transverse identity kernel $\PT(x,x')$ is defined as
\begin{align}
	\PT_{\mu\nu'}(x,x')=g_{\mu\nu'}(x,x')\delta_g(x,x')+\nabla_{\mu}\nabla_{\nu'}\Delta_{sc}(x,x'),
\end{align}
where the invariant delta distribution is normalized by
$\int\!\dd^4x\sqrt{-g(x)}\,\delta_g(x,x')f(x)=f(x')$.
The kernel satisfies $\nabla^{\mu}\PT_{\mu\nu'}(x,x')=0$.
These equations
are definitions of the inverses of the two factors, not additional field
equations.  Their operator partial fraction gives
\begin{equation}
 \boxed{G_P (x,x')=G_0(x,x')-G_M(x,x')}
 \label{eq:partialfraction}
\end{equation}
obeys
\begin{equation}
 \boxed{\Pop G_P (x,x')=i \PT (x,x')}.
 \label{eq:normalization}
\end{equation}
Indeed, $\Mop[G_0(x,x')-G_M(x,x')]=-M^2G_M(x,x')$ and
$\Mop^2[G_0(x,x')-G_M(x,x')]=- M^2 i \PT(x,x')-M^4G_M(x,x')$, which proves
Eq.~\eqref{eq:normalization}.  When contracted with a conserved current,
$\PT$ acts as the identity.

The same subtraction has a useful action-level interpretation.  Introduce an
auxiliary vector $U_\mu$ in Eq.~\eqref{eq:action} and write {Eq.~\eqref{eq:action} as} 
\begin{equation}
 \mathcal L=-\frac14F(A)^2+U_\mu B^\mu(A)+\frac{M^2}{2}U_\mu U^\mu - J_\mu A^\mu.
 \label{eq:auxiliaryL}
\end{equation}
Eliminating $U_\mu=-B_\mu/M^2$ recovers Eq.~\eqref{eq:action}.  Eq.~\eqref{eq:auxiliaryL}, however, still mixes $A_\mu$ and $U_\mu$.  Up to a
boundary term,
\begin{equation}
 U_\mu B^\mu(A)=-\frac12F_{\mu\nu}(U)F^{\mu\nu}(A).
\end{equation}
Consequently, the redefinition $A^0_{\mu}=A_{\mu}+U_{\mu}$, or equivalently $A_{ \mu}=A^0_{ \mu}-U_{ \mu}$,
diagonalizes the two kinetic terms and gives
\begin{equation}
 \mathcal L=-\frac14F(A^0)^2 - J\!\cdot A^0
 +\frac14F(U)^2+\frac{M^2}{2}U^2 + J\!\cdot U.
 \label{eq:diagonalL}
\end{equation}
Therefore, the purpose of passing from Eq.~\eqref{eq:auxiliaryL} to
Eq.~\eqref{eq:diagonalL} is to display the pole content directly:
$A^0_\mu$ is a Maxwell field, whereas $U_\mu$ is an opposite-sign massive
vector.  Their relative sign independently explains the propagator
combination $G_P(x,x')=G_0(x,x')-G_M(x,x')$.  This diagonalization is not required to solve the
classical Green-function problem, but it makes the origin of the subtraction
and the negative-residue massive pole explicit.

\section{Invariant bitensor formalism}

Four-dimensional de Sitter spacetime is represented as the
hyperboloid
\begin{equation}
 \eta_{AB}X^A X^B=H^{-2},
\end{equation}
embedded in five-dimensional Minkowski spacetime.  For two points
$x$ and $x'$, the standard de Sitter-invariant separation variable is
defined by
\begin{equation}
 Z(x,x')
 =
 H^2\eta_{AB}X^A(x)X^B(x').
 \label{eq:embeddingZ}
\end{equation}
For points connected by a spacelike geodesic of length
$\ell(x,x')$, this invariant satisfies
\begin{equation}
 Z(x,x')=\cos \bigl( H\ell(x,x') \bigr).
 \label{eq:Zdistance}
\end{equation}
For timelike or null separation, Eq.~\eqref{eq:embeddingZ} provides
the fundamental definition, while Eq.~\eqref{eq:Zdistance} is
understood through analytic continuation.  In particular,
$Z=1$ for null geodesic, whereas timelike-related points have $Z>1$
within the normal geodesic domain.

It is convenient to introduce the equivalent hypergeometric variables
\begin{equation}
 u(x,x')=\frac{1+Z(x,x')}{2}.
% \qquad
% y(x,x')=2\left[1-Z(x,x')\right]=4(1-u).
 \label{eq:invariantdistance}
\end{equation}
We denote the geodesic distance by $\ell(x,x')$ rather than by
$\mu(x,x')$, because the symbol
\begin{equation}
 \mu\equiv\frac{M^2}{H^2},
 \label{eq:dimensionlessmass}
\end{equation}
will be reserved below for the dimensionless BP mass
parameter.

The de Sitter-invariant vector and bitensor building blocks used in
the propagator decomposition are
\begin{align}
 p_\mu(x,x')&=\nabla_\mu Z(x,x'),  \\
 p_{\nu'}(x,x')&=\nabla_{\nu'}Z(x,x'),  \\
 \qquad
 q_{\mu\nu'}(x,x')
 &=\nabla_\mu\nabla_{\nu'}Z(x,x').
 \label{eq:bitensorbuildingblocks}
\end{align}
Here $\nabla_\mu$ acts at $x$, while $\nabla_{\nu'}$ acts at $x'$.
Thus, $p_\mu$ and $p_{\nu'}$ are covectors at their respective
points, and $q_{\mu\nu'}$ is a mixed rank-$(1,1)$ bitensor.

In four-dimensional spacetime, these objects satisfy
\begin{align}
 p^\mu p_\mu
 &=H^2(1-Z^2),
 &
 \Boxop Z
 &=-4H^2Z,
 \label{eq:Zidentities1}\\
 \nabla_\alpha p_\mu
 &=-H^2Zg_{\alpha\mu},
 &
 p^\alpha q_{\alpha\nu'}
 &=-H^2Zp_{\nu'},
 \label{eq:Zidentities2}\\
 \nabla_\alpha q_{\mu\nu'}
 &=-H^2g_{\alpha\mu}p_{\nu'}.
 \label{eq:Zidentities3}
\end{align}

At separated points, a de Sitter-invariant vector two-point function
can be expanded in the two independent mixed bitensors introduced
above:
\begin{equation}
 G_{\mu\nu'}(x,x')
 =
 A(Z)q_{\mu\nu'}
 +B(Z)p_\mu p_{\nu'}.
 \label{eq:ABansatz}
\end{equation}
Here $A(Z)$ and $B(Z)$ are scalar coefficient functions. Depending on the chosen
Green-function prescription, these coefficients are understood as
the appropriate boundary values or distributions.  Possible contact
terms at $x=x'$ are fixed separately by the inhomogeneous
normalization equation.

To impose transversality at the unprimed point, we take the divergence
of Eq.~\eqref{eq:ABansatz}.  Applying the product rule gives
\begin{align}
 \nabla^\mu G_{\mu\nu'}
 ={}&
 A'(Z)p^\mu q_{\mu\nu'}
 +A(Z)\nabla^\mu q_{\mu\nu'}
 +B'(Z)p^\mu p_\mu p_{\nu'}
 \nonumber\\
 &
 +B(Z)(\Boxop Z)p_{\nu'}
 +B(Z)p^\mu q_{\mu\nu'}.
 \label{eq:ABdivergence}
\end{align}
The identities above imply
$\nabla^\mu q_{\mu\nu'}=-4H^2p_{\nu'}$ and hence
\begin{equation}
 \nabla^\mu G_{\mu\nu'}=H^2p_{\nu'}
 \left[(1-Z^2)B'-5ZB-ZA'-4A\right].
\end{equation}
Note that a prime on a coefficient function denotes an ordinary derivative with respect to $Z$.
Therefore, $\nabla^\mu G_{\mu\nu'}=0$ gives
\begin{equation}
 (1-Z^2)B'-5ZB-ZA'-4A=0.
 \label{eq:transversality}
\end{equation}

At $x\neq x'$, the massive equation Eq.~\eqref{eq:GMdef} is
	\begin{equation}
		(\Box - M^2 - 3 H^2) G_{\mu\nu'}=\nabla_{\mu}\nabla_{\nu'}G_{sc}.
	\end{equation}
In the bitensor decomposition the r.h.s. becomes 
\begin{equation}
\nabla_{\mu}\nabla_{\nu'}G_{sc}=G^{'}_{sc} q_{\mu\nu'}+G^{''}_{sc}p_\mu p_{\nu'}.
\end{equation}
Introduce the master function $\gamma(Z)$ by
\begin{align}
 A&=-\frac{(1-Z^2)\gamma'-3Z\gamma}{3H^2},\label{eq:Afromgamma}\\
 B&=-\frac{Z\gamma'+3\gamma}{3H^2},
 \label{eq:Bfromgamma}
\end{align}
which satisfy Eq.~\eqref{eq:transversality} identically. 
The non-contact massive master function simplifies as
\begin{equation}
 (1-Z^2)\gamma''-6Z\gamma'
 -\left(6+\frac{M^2}{H^2}\right)\gamma=\frac{Z G^{'}_{sc} -(1-Z^2) G^{''}_{sc}}{3}.
 \label{eq:masterodeFull}
\end{equation}
The complete transverse Green function  contains a particular solution
fixed by $G_{sc}$ of the transverse projector and a general solution fixed by the homogeneous part, which are displayed
explicitly below.

\section{Exact transverse propagator}

Here we solve the non-contact massive master function Eq.~\eqref{eq:masterodeFull}. Let
\begin{equation}
 y=2(1-Z),\qquad u=1-\frac y4=\frac{1+Z}{2},
 \label{eq:yuvariables}
\end{equation}
and define
\begin{equation}
 \nu^2=\frac14-\frac{M^2}{H^2},\qquad
 a=\frac52+\nu,\qquad b=\frac52-\nu.
 \label{eq:hyperparameters}
\end{equation}
Since $\partial_Z=\tfrac12\partial_u$ and
$1-Z^2=4u(1-u)$, the homogeneous part of Eq.~\eqref{eq:masterodeFull} becomes
\begin{equation}
 \mathcal L_u\gamma=0,\qquad
 \mathcal L_u\equiv u(1-u)\partial_u^2+(3-6u)\partial_u
 -\left(6+\frac{M^2}{H^2}\right).
 \label{eq:hyperoperator}
\end{equation}
This is the Gauss equation because $a+b=5$, $ab=6+M^2/H^2$,
and $c=3$.  Hence ${}_2F_1(a,b;3;u)$ solves its homogeneous part.

For the full transverse Feynman kernel, however, we need the scalar function $G_{sc}$~\cite{Allen:1987tz,TsamisWoodard2007,Onemli:2002hr}
\begin{align}
 G_{sc}=\frac{H^2}{4\pi^2 y} + \frac{H^2}{8\pi^2} \ln{\frac{a a'}{y}},
\end{align}
where $a=a(\eta), a'=a(\eta')$ are scale factors at the two points $x, x'$ in the flat coordinate $ds^2= a(\eta)^2  (-d\eta^2 + d\vec{x}^2)$.   Substituting it to the r.h.s. of Eq.~\eqref{eq:masterodeFull} and by
\begin{equation}
\Box \ln a = -3 H^2,
\end{equation}
the projector contribution gives
\begin{equation}
 \boxed{\mathcal L_u\gamma_M
 =-\frac{3H^2}{32\pi^2}\frac{1}{(1-u)^2}.}
 \label{eq:inhomogeneousmaster}
\end{equation}
Using
\begin{equation}
 \mathcal L_u(1-u)^{-2}
 =-\frac{M^2}{H^2}(1-u)^{-2},
\end{equation}
we obtain a particular solution  
\begin{equation}
 \gamma_{\rm part}=
 \frac{3H^4}{32\pi^2M^2}(1-u)^{-2}.
\end{equation}
Hence, the solution regular at $u=0$ is the particular
solution plus $C_M\; {}_2F_1(a,b;3;u)$. To determine the constant $C_M$, analytic continuation about $u=1$
shows that
\begin{equation}
 {}_2F_1(a,b;3;u)\sim
 \frac{2}{\Gamma(a)\Gamma(b)}(1-u)^{-2}+\cdots .
\end{equation}
Cancellation of this overly singular double pole fixes the relative
coefficient, while the Feynman--Hadamard contact normalization  fixes the
overall factor.  The resulting normalized massive transverse master function
is
\begin{equation}
 \gamma_M=
 \frac{3H^4}{32\pi^2M^2}
 \left[\frac1{(1-u)^2}
 -\frac{\Gamma(a)\Gamma(b)}2
 {}_2F_1(a,b;3;u)\right].
 \label{eq:gammamassive}
\end{equation}

The finite limit $M\to 0$  of $\gamma_M$ will give the massless transverse master function $\bar\gamma$.  Put $\mu=M^2/H^2$.  Then the parameters Eq.~\eqref{eq:hyperparameters} become
 \begin{align}
  \nu&=\frac12-\mu+O(\mu^2),\\
  a&=3-\mu+O(\mu^2), \\
  b&=2+\mu+O(\mu^2).
 \end{align}
The first-order expansion of ${}_2F_1$ is
\begin{align}
 \frac{\Gamma(a)\Gamma(b)}2{}_2F_1(a,b;3;u)
 ={}&\frac{1}{(1-u)^2} \nonumber \\
 &-\mu\left[\frac{1}{1-u}+\frac{\ln(1-u)}{u^2}
 \right] \nonumber \\
 &-\frac{\mu}{u}+O(\mu^2).
\end{align}
Using $1-u=y/4$, the massless limit $\mu \to 0$ gives
\begin{equation}
 \bar\gamma=\frac{3H^2}{32\pi^2}
 \left[\frac4{y}
 +\frac{\ln(y/4)}{(1-y/4)^2}
 +\frac1{1-y/4}\right].
 \label{eq:gammamassless}
\end{equation}
The exact BP master function is
\begin{equation}
 \boxed{\gamma_P^F(y_F)=\bar\gamma^F(y_F)-\gamma_M^F(y_F)},
 \label{eq:gammaP}
\end{equation}
where the superscript $F$ indicates the Feynman prescription $y_F=y+\ii0$ (or equivalently $u_F=u-\ii0$). 
The $A_P$ and $B_P$ can be reconstructed from
Eqs.~\eqref{eq:Afromgamma}--\eqref{eq:Bfromgamma}.

The subtraction in Eq.~\eqref{eq:gammaP} is inherited directly from
$G_P(x,x')=G_0(x,x')-G_M(x,x')$.  Because Eqs.~\eqref{eq:Afromgamma}--\eqref{eq:Bfromgamma}
are linear in $\gamma$, the same subtraction applies to the master functions.
In particular, the common leading short-distance singularities of the two
second-order sectors cancel in $\gamma_P^F$, producing the improved ultraviolet
behavior analyzed in the next section.

For completeness, the potential propagator is defined only up to a
longitudinal addition
\begin{equation}
 \Delta_{\mu\nu'}\longrightarrow \Delta_{\mu\nu'}
 +\nabla_\mu\nabla_{\nu'}S(x,x').
 \label{eq:longitudinaladdition}
\end{equation}
Its contribution to the response of a conserved current is
\begin{align}
 \delta A_\mu(x)
 & =\int\!\dd^4x'\sqrt{-g'}\,
 \nabla_\mu\nabla_{\nu'}S(x,x')J^{\nu'}(x')\\
 &=-\nabla_\mu\int\!\dd^4x'\sqrt{-g'}\,
 S(x,x')\nabla_{\nu'}J^{\nu'}(x'),
\end{align}
up to a boundary term.  Therefore, it vanishes for a conserved source under the
standard Green-function boundary conditions; more generally it changes the
potential only by a pure gauge contribution.  A curl removes the ambiguity
locally because
$\nabla_{[\rho}\nabla_{\mu]}\nabla_{\nu'}S=0$, and a primed curl does the
same at the other point.  Finally, if $S=f(x)+g(x')$, then
$\nabla_\mu\nabla_{\nu'}S=0$ even before taking a curl.  This includes the
usual separated de Sitter-breaking ambiguity and explains why it cannot affect
the one- or two-curl electromagnetic kernels.

\section{Retarded kernel and light-cone structure}
\label{sec:retarded}

The Feynman prescription $y_F=y+\ii0$, equivalently $u_F=u-\ii0$,
specifies boundary values on the two sides of the timelike branch cut.  Their
discontinuity is the causal spectral kernel,
\begin{equation}
 \operatorname{Disc}\gamma_P^F
 =\gamma_P^F(u+\ii0)-\gamma_P^F(u-\ii0)
 =2\ii\,\operatorname{Im}\gamma_P^F(u+\ii0).
\end{equation}
Multiplication by $\theta(t-t')$ selects the future-supported part, that is, the retarded Green's function
$G^{\ret}=i\Delta^{\ret}=2i\theta(t-t')\operatorname{Im} G_F $. Then the retarded master function $\gamma^{\ret}$ in terms of $\Delta^{\ret}$  is
\begin{equation}
 \gamma_P^{\ret}=2i\theta(t-t')\,\operatorname{Im}\gamma_P^F.
 \label{eq:retprescription}
\end{equation}
The step function supplies retarded rather than
advanced support.

This prescription has an invariant interpretation in the maximally analytic
de Sitter framework of Refs.~\cite{BrosGazeauMoschella1994,BrosMoschella1996,Moschella2024}.
There the two Wightman functions are opposite boundary values of a function
holomorphic on the complexified de Sitter manifold.  Their difference is the
commutator kernel, and multiplication by the invariant future-support
function $\theta_+(x,x')$ gives the retarded Green function.  In the present
classical calculation we use this boundary-value statement only to fix the
causal distribution; it does not impose a quantum Lee--Wick contour.  Locally,
the mechanism is visible from
\begin{equation}
 \frac1{y\pm\ii0}=\operatorname{PV}\frac1y\mp\ii\pi\delta(y),
 \qquad
 \ln(y\pm\ii0)=\ln|y|\pm\ii\pi\theta(-y),
 \label{eq:boundaryvaluedistributions}
\end{equation}
with the logarithmic sign understood on the timelike side according to our
choice of $y$.  Thus, poles generate distributions supported on the null cone,
whereas the hypergeometric or logarithmic cut generates a timelike tail.  The
bitensors $q_{\mu\nu'}$ and $p_\mu p_{\nu'}$ carry the vector index structure
without changing this scalar support analysis.

To audit the short-distance coefficients, put $x=1-u=y/4$ and use Eq.~\eqref{eq:dimensionlessmass}.
%$\mu=M^2/H^2$.
If
\begin{equation}
 f(x)=\frac{\Gamma(a)\Gamma(b)}2{}_2F_1(a,b;3;1-x),
\end{equation}
then Eq.~\eqref{eq:hyperoperator} becomes
\begin{equation}
 x(1-x)f''+(3-6x)f'-(6+\mu)f=0.
\end{equation}
Since $a+b-c=2$, its Frobenius expansion about the light cone $x=0$
contains a double pole, a simple pole, and a logarithm.  Substitution of
\begin{equation}
 f(x)=x^{-2}+c_{-1}x^{-1}+c_{\log}\ln x+O(1)
\end{equation}
fixes $c_{-1}=-\mu$ and
$c_{\log}=-\mu(2+\mu)/2$.  Hence,
\begin{align}
 \frac{\Gamma(a)\Gamma(b)}2{}_2F_1(a,b;3;1-x)
 ={}&\frac1{x^2}-\frac{\mu}{x}\nonumber\\
 &-\frac{\mu(2+\mu)}2\ln x+O(1).
 \label{eq:hypernearcone}
\end{align}
Thus, Eq.~\eqref{eq:hypernearcone} is the near-cone expansion of the
hypergeometric term in Eq.~\eqref{eq:gammamassive}, not a separate ansatz.
It also can be obtained by applying the logarithmic expansion of ${}_2F_1$, see Eq.(15.8.11) and Eq.(15.8.12) of~\cite{NIST:DLMF}. 
Inserting it into Eq.~\eqref{eq:gammamassive} cancels the explicit $x^{-2}$
term and gives the first line below; directly expanding
Eq.~\eqref{eq:gammamassless} gives the second:
\begin{align}
 \gamma_M^F&=\frac{3H^2}{8\pi^2y_F}
+\frac{3H^2}{64\pi^2}(2+\mu)\ln\frac{y_F}{4}+O(1),
 \label{eq:massivenearcone}\\
 \bar\gamma_F&=\frac{3H^2}{8\pi^2y_F}
+\frac{3H^2}{32\pi^2}\ln\frac{y_F}{4}+O(1).
 \label{eq:maxwellnearcone}
\end{align}
Eqs.~\eqref{eq:massivenearcone} and
\eqref{eq:maxwellnearcone} make the key cancellation explicit: the Maxwell
and massive-vector sectors have exactly the same strongest short-distance
singularity, $3H^2/(8\pi^2y_F)$.  Since the BP kernel is their
difference, this common pole cancels and
\begin{equation}
\gamma_P^F=-\frac{3M^2}{64\pi^2}
 \ln\frac{y_F}{4}+O(1).
 \label{eq:logsoftening}
\end{equation}
Keeping the next logarithmic coefficient gives
\begin{equation}
  \boxed{\gamma_P^{\ret}=-i \theta_+\theta(-y)
 \left[g_0+g_1y+O(y^2)\right],}
 \label{eq:gammaretseries}
\end{equation}
where
\begin{equation}
 g_0=\frac{3M^2}{32\pi},\qquad
 g_1=\frac{M^2}{128\pi}(8+\mu).
 \label{eq:g0g1}
\end{equation}
Using Eqs.~\eqref{eq:Afromgamma}--\eqref{eq:Bfromgamma} distributionally,
\begin{align}
i A_P^{\ret}&=\theta_+\theta(-y)\frac{3M^2}{32\pi H^2}+O(y),\\
i B_P^{\ret}&=-\theta_+\frac{M^2}{16\pi H^2}\delta(y)
 +\theta_+\theta(-y)\frac{M^2(\mu-10)}{192\pi H^2}+O(y).
\end{align}
Substitution into Eq.~\eqref{eq:ABansatz} displays the bitensors and causal
distributions together:
\begin{equation}
 \begin{aligned}
 i G_{\mu\nu'}^{P,\ret}={}&
 -\theta_+\frac{M^2}{16\pi H^2}
 \delta(y)p_\mu p_{\nu'}\\
 &+\theta_+\theta(-y)
 \frac{3M^2}{32\pi H^2}q_{\mu\nu'}\\
 &+\theta_+\theta(-y)
 \frac{M^2(\mu-10)}{192\pi H^2}
 p_\mu p_{\nu'}+O(y).
 \end{aligned}
 \label{eq:nearconevectorpropagator}
\end{equation}
%Here $q_{\mu\nu'}$ and $p_\mu p_{\nu'}$ are the two invariant vector bitensors.
Eq.~\eqref{eq:nearconevectorpropagator} is the vector analogue
of the familiar graviton representation as invariant rank-$(2,2)$ bitensors
multiplied by null-cone delta distributions and an interior Heaviside tail.
The distributions have a direct causal interpretation.  A factor
$\delta(y)$ is supported exactly on the null cone, whereas $\theta(-y)$ is
supported at timelike separation inside it.  The massless and massive
second-order propagators possess the common pole identified above; Therefore, its
associated strongest wavefront cancels in $G_P=G_0-G_M$.
Nevertheless,
$\gamma_P^{\ret}$ still has a step across $y=0$.  Differentiating this step in
Eqs.~\eqref{eq:Afromgamma}--\eqref{eq:Bfromgamma} produces the weaker
$\delta(y)p_\mu p_{\nu'}$ term in $B_P^{\ret}$, while the accompanying
$\theta(-y)$ terms describe an interior tail.  Thus, the potential is
softened but not smooth across the light cone.  Moreover, this conclusion is
about the gauge-dependent potential kernel; differentiation can reorganize
its singularities, so the physical field-strength wavefront must be examined
after taking a curl.

\section{Gauge-invariant electromagnetic kernels}

The potential response to a conserved current is
\begin{equation}
 A_\mu(x)=-i \int\!\dd^4x'\sqrt{-g'}\,
 G_{\mu\rho'}(x,x')J^{\rho'}(x').
\end{equation}
Because $A_\mu$ is gauge dependent, the physical response is instead
$F_{\mu\nu}=2\nabla_{[\mu}A_{\nu]}$.  This motivates defining the one-curl
kernel: it maps the current directly to the observable field strength and,
as shown after Eq.~\eqref{eq:longitudinaladdition}, annihilates every
longitudinal addition $\nabla_\mu\nabla_{\rho'}S$.  One curl of
Eq.~\eqref{eq:ABansatz} yields
\begin{equation}
 \mathcal K_{\mu\nu|\rho'}
 \equiv2\nabla_{[\mu}G_{\nu]\rho'}
 =\mathcal C_F(Z)
 (p_\mu q_{\nu\rho'}-p_\nu q_{\mu\rho'}),
 \label{eq:onecurl}
\end{equation}
so that
\begin{equation}
 F_{\mu\nu}(x)=-i \int\!\dd^4x'\sqrt{-g'}\,
 \mathcal K_{\mu\nu|\rho'}(x,x')J^{\rho'}(x').
\end{equation}
One curl is sufficient for the response to a conserved current.  A second,
primed curl would instead construct the field-strength--field-strength kernel,
which is gauge invariant at both points.  The scalar coefficient in
Eq.~\eqref{eq:onecurl} is
\begin{equation}
 \mathcal C_F=A'-B=\frac1{3H^2}
 \left[-(1-Z^2)\gamma''+6Z\gamma'+6\gamma\right].
 \label{eq:CFoperator}
\end{equation}
In $y$ variables,
\begin{equation}
 \mathcal C_F=\frac1{3H^2}
 \left[-(4y-y^2)\partial_y^2-(12-6y)\partial_y+6\right]\gamma.
\end{equation}
Applying this operator to Eq.~\eqref{eq:gammaretseries} gives the audited result
\begin{equation}
 \boxed{\mathcal C_F^{P,\ret}=-i\theta_+
 \left[\frac{M^2}{4\pi H^2}\delta(y)
 -\frac{M^2(2+\mu)}{32\pi H^2}\theta(-y)+O(y)\right]}.
 \label{eq:CFnearcone}
\end{equation}
No derivative of a delta function occurs after one curl.

The local covariant limit checks Eq.~\eqref{eq:CFnearcone}.  With Synge's
world function $\sigma$,
\begin{align}
 y&=2H^2\sigma+O(\sigma^2), \\
 p_\mu &=-H^2\sigma_{;\mu}+\cdots,\\
 q_{\mu\nu'}&=H^2g_{\mu\nu'}+\cdots.
\end{align}
Consequently Eq.~\eqref{eq:onecurl} becomes
\begin{align}
i \mathcal K^{P,\ret}_{\mu\nu|\nu'}={}&
 -\frac{M^2}{4\pi}\theta_+\delta(\sigma)
 \sigma_{;[\mu}g_{\nu]\nu'}\nonumber\\
 &+\frac{M^4+2M^2H^2}{16\pi}\theta_+\theta(-\sigma)
 \sigma_{;[\mu}g_{\nu]\nu'}+\cdots.
\end{align}
The flat limit $H\to 0$ of this local expansion agrees with the local expansion in Minkowski spacetime Eq.~\eqref{eq:flatKmunu}.

For the full timelike tail let $u>1$ and write
\begin{equation}
 \gamma_{P,\mathrm{tail}}^{\ret}= -i \theta_+\theta(u-1)\Gamma_P(u).
\end{equation}
Define $\alpha=1/2-\nu$, $\beta=1/2+\nu$, and
\begin{multline}
 \mathcal D_\nu(u)=\frac{4\sin(\pi\alpha)}{\Gamma(\beta)\Gamma(3-\beta)}
 \frac1{(u-1)^2}\\
 \times\int_{1/u}^{1}\!\dd t\,
 t^{\beta-1}(1-t)^{2-\beta}(ut-1)^{-\alpha}.
 \label{eq:Dnu}
\end{multline}
we obtain 
\begin{equation}
 \Gamma_P(u)=-\frac{3H^2}{16\pi u^2}
 +\frac{3H^4\Gamma(a)\Gamma(b)}{64\pi^2M^2}\mathcal D_\nu(u).
 \label{eq:fulltail}
\end{equation}
With the prescription inherited from $y_F=y+i0$, one has $u_F=u-i0$ and
\begin{equation}
 \mathcal D_\nu(u)=-2\,\operatorname{Im}{}_2F_1(a,b;3;u-i0).
 \label{eq:Ddiscontinuity}
\end{equation}
In particular,  in the limit $u\to 1^{+}$ we obtain,
\begin{equation}
 \mathcal D_\nu(1^+)=
 \frac{2\pi\mu(2+\mu)}{\Gamma(a)\Gamma(b)},
 \label{eq:Dnucone}
\end{equation}
which gives $\Gamma_P(1^+)=3M^2/(32\pi)$.  Appendix~\ref{app:branchcut}
derives Eqs.~\eqref{eq:Dnu}--\eqref{eq:Dnucone}, including their normalization
and sign, directly from Euler's integral.
The first term is the smooth Maxwell master tail.  In terms of $u$, the curl operator is
\begin{equation}
 \Lie_F=u(u-1)\frac{\dd^2}{\dd u^2}
 +3(2u-1)\frac{\dd}{\dd u}+6.
\end{equation}
It satisfies $\Lie_Fu^{-2}=0$.  Hence, the smooth Maxwell master tail is
annihilated by the curl, and the exact physical interior coefficient is
\begin{equation}
 \boxed{\mathcal C_{F,\mathrm{tail}}^P(u)= -i \theta_+ \theta(u-1)
 \frac{H^2\Gamma(a)\Gamma(b)}{64\pi^2M^2}
 \Lie_F\mathcal D_\nu(u)}.
 \label{eq:exactphysicaltail}
\end{equation}
Thus, the gauge-invariant interior electromagnetic tail is entirely massive-sector generated.

A second curl can be expressed using
\begin{align}
 \mathbb P_{\mu\nu|\rho'\sigma'}={}&p_{\rho'}(p_\mu q_{\nu\sigma'}-p_\nu q_{\mu\sigma'})
 -(\rho'\leftrightarrow\sigma'),\\
 \mathbb Q_{\mu\nu|\rho'\sigma'}={}&q_{\mu\rho'}q_{\nu\sigma'}-q_{\mu\sigma'}q_{\nu\rho'}.
\end{align}
The doubly gauge-invariant kernel is
\begin{equation}
 \mathcal H_{\mu\nu|\rho'\sigma'}
 =\mathcal C_F'(Z)\mathbb P_{\mu\nu|\rho'\sigma'}
 +2\mathcal C_F(Z)\mathbb Q_{\mu\nu|\rho'\sigma'},
\end{equation}
and satisfies the Bianchi identities at both endpoints.

\section{Flat and static limits}

\subsection{Minkowski spacetime}
Recall the Klein-Gordon propagator in the 4d Minkowski spacetime. The spatial Fourier transformation reduces the massless and
massive second-order retarded equations to forced oscillators with frequencies
$k$ and $\omega_k=\sqrt{k^2+M^2}$.  Their normalized retarded solutions are
$\theta(T)\sin(kT)/k$ and
$\theta(T)\sin(\omega_kT)/\omega_k$, respectively. Therefore, applying
$G_P=G_0-G_M$ mode by mode gives
\begin{equation}
 g_{P,k}^{\ret}(T)= -i \theta(T)\left[
 \frac{\sin(kT)}k-\frac{\sin(\sqrt{k^2+M^2}\,T)}{\sqrt{k^2+M^2}}\right].
 \label{eq:flatmodekernel}
\end{equation}

The standard three-dimensional inverse transforms of the massless and massive
retarded kernels contain the same null-cone delta distribution, which cancels
in their difference.  The remaining massive interior term gives
\begin{equation}
 \boxed{G_{P,\mathrm{flat}}^{\ret}(T,R)= -i
 \theta(T)\theta(T-R)\frac{M J_1(Ms)}{4\pi s}},
 \label{eq:flatbessel}
\end{equation}
where $s=\sqrt{T^2-R^2}$.
This is finite at the cone, $G_P^{\ret}\to -i M^2/(8\pi)$ as $s\to0^+$. 

With the transverse projector, the retarded Green's function is
\begin{align}
G^{\mathrm{flat,\ret}}_{P,\mu\nu} = \left(\eta_{\mu\nu}-\frac{\partial_\mu \partial_\nu}{\Box}\right) G_{P,\mathrm{flat}}^{\ret}.
\end{align}
Therefore, the flat version of the electromagnetic kernel Eq.~\eqref{eq:onecurl} is
\begin{equation}
	\mathcal K^{flat,\ret}_{P,\mu\nu|\rho}=2 \partial_{[\mu}\sigma \eta_{\nu]\rho} \frac{d G_{P,\mathrm{flat}}^{\ret}}{d\sigma}.
\end{equation}
With the series expansion of Bessel functions, the local expansion gives
\begin{align}
	\label{eq:flatKmunu}
	i \mathcal K^{flat,\ret}_{P,\mu\nu|\rho}&=\theta_+\Big(-\frac{M^2}{4\pi}\delta(\sigma)+\frac{M^4}{16\pi}\theta(-\sigma)\Big)\sigma_{;[\mu}\eta_{\nu]\rho}+\ldots.
\end{align}

\subsection{de Sitter spacetime}

%For the static patch,
 The metric of de Sitter spacetime in the static patch is given by 
\begin{equation}
 \dd s^2=-(1-H^2r^2)\dd t^2
 +\frac{\dd r^2}{1-H^2r^2}+r^2\dd\Omega_2^2,
 \label{eq:staticmetric}
\end{equation}
with the gauge field $A_{ \mu}=\Phi(r)\dd t$.  Away from the charge, the Maxwell equation
$\nabla_\mu F^{\mu t}=(\Box - 3H^2) A^{t}=0$ reduces to
\begin{equation}
 \frac{-1}{r^2}\frac{\dd}{\dd r}\left(r^2\Phi_0'\right)=0,
\end{equation}
so the general solution is $\Phi_0=C_1+C_2/r$.  Coulomb normalization fixes
$C_2=q/(4\pi)$, and the horizon condition $\Phi_0(H^{-1})=0$ fixes
$C_1=-qH/(4\pi)$.  Hence
\begin{equation}
 \Phi_0=\frac{q}{4\pi}\left(\frac1r-H\right).
 \label{eq:staticmaxwell}
\end{equation}

The massive equation is
\begin{equation}
 \Phi_M''+\frac2r\Phi_M'-\frac{M^2}{1-H^2r^2}\Phi_M=0.
\end{equation}
As $H\to0$ the horizon
moves to infinity and this massive equation reduces to the
flat radial Proca equation, whose solution is  $\Phi^{flat}_M=q e^{-Mr}/(4\pi r)$ following the Coulomb normalization and decay at the limiting
horizon. So the flat solution is 
$\Phi^{flat}_P=q (1-e^{-Mr})/(4\pi r)$, consistent with the time integral of Eq.~\eqref{eq:flatbessel}. 

Let
\begin{equation}
 \alpha_s=\frac{1+\sqrt{1-4\mu}}4,\qquad
 \beta_s=\frac{1-\sqrt{1-4\mu}}4,
\end{equation}
and
\begin{equation}
 C_{\alpha_s\beta_s}=
 \frac{\sqrt\pi}{\Gamma(3/2-\alpha_s)\Gamma(3/2-\beta_s)}.
\end{equation}
The horizon-regular, Coulomb-normalized massive solution is
\begin{multline}
 \Phi_M=\frac{qH}{4\pi C_{\alpha_s\beta_s}}(1-H^2r^2)\\
 \times{}_2F_1\left(\frac32-\alpha_s,\frac32-\beta_s;2;1-H^2r^2\right).
 \label{eq:staticmassive}
\end{multline}
Therefore, Eqs.~\eqref{eq:staticmaxwell} and \eqref{eq:staticmassive} give
\begin{equation}
 \boxed{\Phi_P=\Phi_0-\Phi_M}.
\end{equation}

Near the charge $r\to 0$,
\begin{equation}
 \Phi_M=\frac{q}{4\pi r}+\Phi_M^{(0)}+\frac{qM^2}{8\pi}r+O(r^2),
 \label{eq:staticnearorigin}
\end{equation}
then 
\begin{equation}
\Phi_P=\Phi_P(0) - \frac{q M^2}{8\pi}r, 
\end{equation}
with the exact horizon-referenced central value
\begin{equation}
	\Phi_P(0)=\frac{qH}{4\pi}\left[
	2\frac{\Gamma(3/2-\alpha_s)\Gamma(3/2-\beta_s)}
	{\Gamma(1/2+\alpha_s)\Gamma(1/2+\beta_s)}-1\right].
	\label{eq:staticcentral}
\end{equation}
So $\Phi_P(0)$ is finite and the electric field  is
\begin{equation}
	\label{eq:fieldEr}
 \boxed{E_{\hat r}(0)=\frac{qM^2}{8\pi}}.
\end{equation}

\begin{figure}[!htbp]
	\centering
	\includegraphics[width=\linewidth]{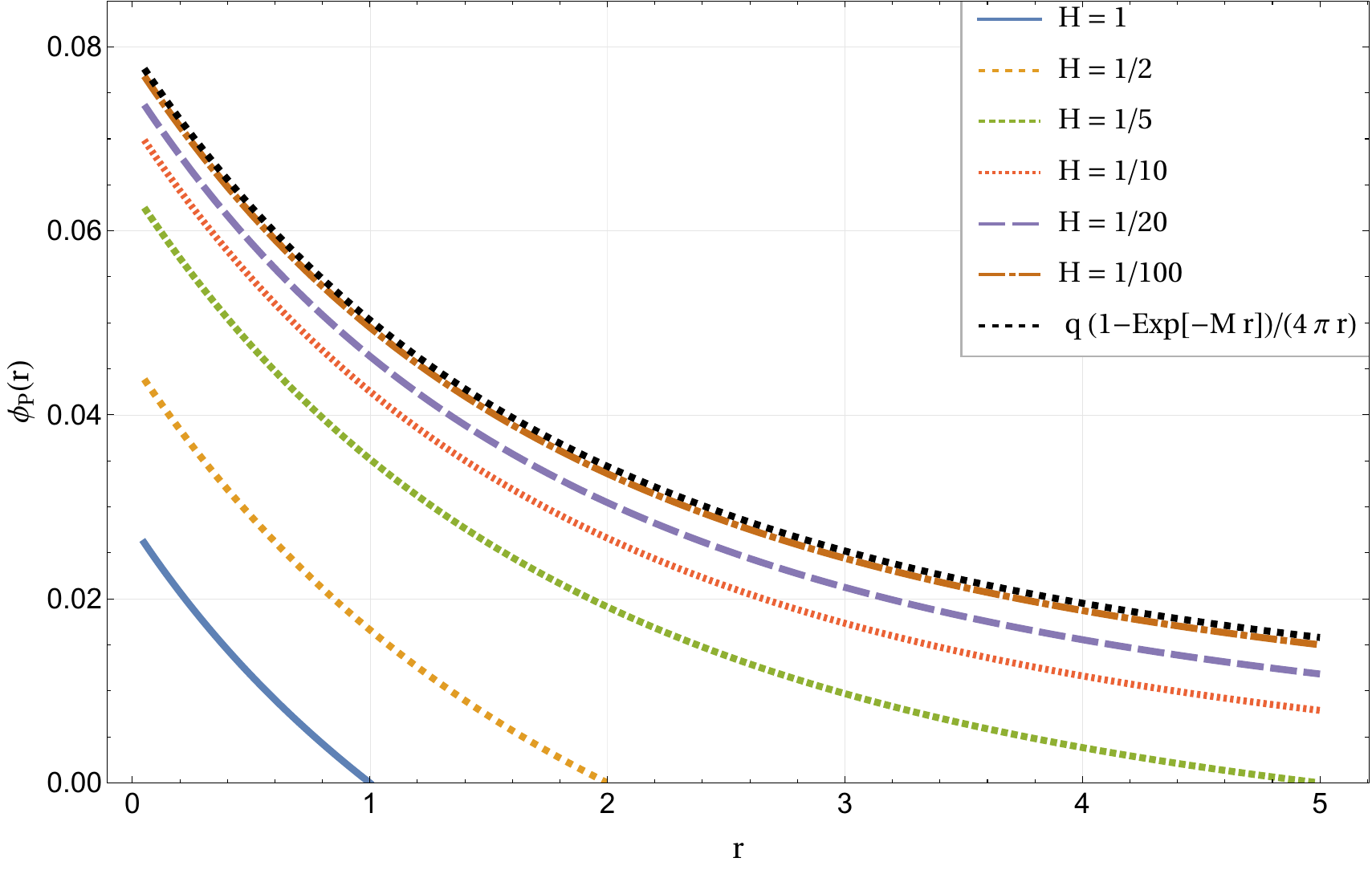}
	\caption{$\Phi_P(r)$ and $\Phi^{flat}_P(r)$ with $q=1,M=1$. }
	\label{fig:numericalFlat}
\end{figure}

As $H\to0$, this series reduces to 
\begin{equation}
\boxed{\Phi_P(r)\longrightarrow\frac{q}{4\pi}(M - M^2 r + O(r^2))},
\end{equation}
which agrees with the series of the flat solution $\Phi^{flat}_P$
\begin{equation}
 \frac{q}{4\pi r}(1-e^{-Mr})\longrightarrow \frac{q}{4\pi}\left(M - M^2 r+ O(r^2)\right)
 \label{eq:staticflatlimit}
\end{equation}
as expected.  A numerical comparison is given further for the flat limit. The numerical value of $\Phi_P(r)$ with $q=M=1$ is given in FIG.~\ref{fig:numericalFlat} for various values of $H$. As $H$ goes to $0$, it approaches the flat one $\Phi^{flat}_P(r)$ as expected.
Thus, the analytic near-origin expansion has the correct flat limit.  The
normalization and limits in Eqs.~\eqref{eq:staticmassive} and
\eqref{eq:staticnearorigin}--\eqref{eq:staticflatlimit} are derived in
Appendix~\ref{app:staticnormalization}.

\section{Generalized Gauss law and global completion}

Eq.~\eqref{eq:eom} can be written
\begin{equation}
 \nabla_\mu\mathfrak F^{\mu\nu}=J^\nu,
 \label{eq:generalizedgauss}
\end{equation}
where
\begin{equation}
 \mathfrak F^{\mu\nu}=F^{\mu\nu}
 -\frac1{M^2}(\nabla^\mu B^\nu-\nabla^\nu B^\mu).
\end{equation}
For the factorized static solution, the Maxwell sector obeys $B_0^\mu=0$
away from the source and the massive sector obeys
$B_M^\mu=M^2A_M^\mu$.  Since $A_P=A_0-A_M$, one has
$B_P^\mu=-M^2A_M^\mu$, and therefore
\begin{equation}
 \mathfrak F_P^{\mu\nu}
 =F_0^{\mu\nu}-F_M^{\mu\nu}
 -\frac1{M^2}(-M^2F_M^{\mu\nu})=F_0^{\mu\nu}.
\end{equation}
For $A_0=\Phi_0(r)\dd t$, Eq.~\eqref{eq:staticmetric} gives
$F_0^{rt}=-\Phi_0'$.  Using Eq.~\eqref{eq:staticmaxwell}, the generalized electric field
\begin{equation}
 \mathfrak{E}_{\hat{r}}=\mathfrak F_P^{rt}=-\Phi_0'=\frac{q}{4\pi r^2}.
 \label{eq:staticgeneralizedflux}
\end{equation}
Thus, the generalized flux equals $q$ on every surrounding sphere, although the ordinary electric flux   vanishes as $r\to0$ by  Eq.~\eqref{eq:fieldEr}.  The higher-derivative term carries the missing local flux.

Global de Sitter spacetime has compact $S^3$ spatial slices.  Let $\Sigma$ be
a complete slice with unit normal $n_\nu$.  Integrating
Eq.~\eqref{eq:generalizedgauss} and applying the covariant divergence theorem
gives
\begin{equation}
 Q_{\rm total}=\int_\Sigma\!\dd\Sigma\,n_\nu J^\nu
 =\int_\Sigma\!\dd\Sigma\,n_\nu\nabla_\mu\mathfrak F^{\mu\nu}
 =\int_{\partial\Sigma}\!\dd S_{\mu\nu}\,\mathfrak F^{\mu\nu}.
\end{equation}
Since a global $S^3$ has no boundary, $\partial\Sigma=\varnothing$, and hence
\begin{equation}
 \boxed{Q_{\mathrm{total}}=0}.
 \label{eq:globalneutrality}
\end{equation}
Consequently, a globally admissible completion of a static-patch charge is an antipodal pair $+q$ at $\chi=0$ and $-q$ at $\chi=\pi$.  In global coordinates
\begin{equation}
 \dd s^2=H^{-2}\left[-\dd T^2+\cosh^2T
 (\dd\chi^2+\sin^2\chi\,\dd\Omega_2^2)\right],
\end{equation}
the static/global transformation is
\begin{equation}
 r=H^{-1}\cosh T\sin\chi,\qquad
 \tanh(Ht_s)=\frac{\tanh T}{\cos\chi}.
 \label{eq:staticglobalmap}
\end{equation}
To transform the one-form, write
$\dd t_s=(\partial_Tt_s)\dd T+(\partial_\chi t_s)\dd\chi$.
Differentiating Eq.~\eqref{eq:staticglobalmap} and then setting $T=0$ gives
\begin{equation}
 \begin{aligned}
  r\big|_{T=0}&=H^{-1}\sin\chi,\\
  \left.\frac{\partial t_s}{\partial T}\right|_{T=0}
  &=\frac1{H\cos\chi},\\
  \left.\frac{\partial t_s}{\partial\chi}\right|_{T=0}&=0.
 \end{aligned}
\end{equation}
Consequently a static sector $A^{(s)}=\Phi_s(r)\dd t_s$ induces
\begin{equation}
 A_T^{(s)}=\frac{\Phi_s(\sin\chi/H)}{H\cos\chi},\qquad A_\chi^{(s)}=0.
 \label{eq:globalinitialoneform}
\end{equation}
Therefore, the vanishing of $A_\chi^{(s)}$ is a property of the time-symmetric
slice, rather than an additional gauge choice.  Because each horizon-regular
potential behaves as $\Phi_s=O(\cos^2\chi)$ at the equator, the factor
$1/\cos\chi$ in Eq.~\eqref{eq:globalinitialoneform} is harmless and the north
and south data match smoothly.

It remains to specify data appropriate to the fourth-order equation.  The
factorization does this without introducing independent higher-time-derivative
data by decomposing
$A_P=A_0-A_M$.  Away from contact sources the two sectors satisfy
$\Mop A_0=0$ and $(\Mop-M^2)A_M=0$, so
\begin{equation}
 \Mop A_P=\Mop A_0-\Mop A_M=0-M^2A_M.
\end{equation}
Thus,
\begin{equation}
 \Mop A_P=-M^2A_M.
 \label{eq:factorizedinitialdata}
\end{equation}
Eq.~\eqref{eq:factorizedinitialdata} is included to state the compatible
auxiliary initial datum carried by the massive sector; it is not a new field
equation.  Together with the initial values of $A_0$ and $A_M$ and their first
normal derivatives, it supplies the data required by the factorized
fourth-order Cauchy problem.

This use of initial data follows the prescription for fields on de Sitter spacetime with spacelike past infinity
\cite{HiguchiLee2008}; its electromagnetic implementation in general covariant
gauge was subsequently given in Ref.~\cite{HiguchiLeeNicholas2009}.  A retarded
source integral alone need not reproduce a field already present on an initial
Cauchy surface $\Sigma$.  For each factor $s=0,M$, Green's identity gives the
decomposition
\begin{equation}
 A_{s\mu}(x)=A^{\rm src}_{s\mu}(x)+A^{\rm init}_{s\mu}(x),
 \label{eq:sourceinitialsplit}
\end{equation}
where, in the transverse second-order representation,
\begin{align}
 A^{\rm src}_{s\mu}(x)
 &=\int_{D^+(\Sigma)}\!\dd^4x'\sqrt{-g'}\,
 \Delta^{\ret}_{s\mu\nu'}(x,x')J_s^{\nu'}(x'),
 \label{eq:sourcepiece}\\
 A^{\rm init}_{s\mu}(x)
 &=\int_\Sigma\!\dd\Sigma_{\alpha'}
 \left[\Delta^{\ret}_{s\mu\nu'}\nabla^{\alpha'}A_s^{\nu'}
 -A_s^{\nu'}\nabla^{\alpha'}\Delta^{\ret}_{s\mu\nu'}\right].
 \label{eq:initialpiece}
\end{align}
The orientation of $\dd\Sigma_{\alpha'}$ fixes the equivalent overall sign of
the surface term.  Gauge-fixed formulations replace the displayed normal
derivatives by the corresponding canonical momenta, without changing the
source-plus-initial-data structure.

Applying Eqs.~\eqref{eq:sourceinitialsplit}--\eqref{eq:initialpiece} separately
to the Maxwell and massive factors gives
\begin{equation}
 A_P^{\rm src}=A_0^{\rm src}-A_M^{\rm src},\qquad
 A_P^{\rm init}=A_0^{\rm init}-A_M^{\rm init}.
 \label{eq:factorizedsourceinitial}
\end{equation}
The data in Eq.~\eqref{eq:globalinitialoneform}, together with
Eq.~\eqref{eq:factorizedinitialdata}, specify the two surface terms.  The sum
of the retarded source field and propagated initial field then equals the
static antipodal solution by uniqueness of the factorized Cauchy problem.

\section{ Closing remarks }

The pole decomposition
\begin{equation}
 \frac1{\lambda(1-\lambda/M^2)}=
 \frac1\lambda-\frac1{\lambda-M^2}
\end{equation}
contains an opposite-residue massive pole.  Formally, $G_P=G_0-G_M$ resembles
a two-point spectral sum concentrated at a massless and a massive pole.  It is,
however, a \emph{signed} sum: the relative minus sign is incompatible with an
ordinary positive K\"all\'en--Lehmann measure of the type developed for
positive-metric de Sitter fields in Ref.~\cite{BrosMoschella1996}.  The same
sign produces the $p^{-4}$ ultraviolet cancellation but requires additional
prescriptions in a quantum Lee--Wick interpretation.  Our results concern the
classical retarded kernel, for which both factors are assigned retarded
boundary conditions.

The transverse representation may contain a scalar Green function through the source projector.  Its constant zero mode and separated terms are killed by mixed derivatives, while longitudinal changes alter the response to a conserved source only by a gauge transformation. Therefore, we claim a de Sitter-covariant classical transverse response and gauge-invariant field-strength kernels, not a preferred quantum Feynman vacuum in every gauge.

Finally, alternative curved-space covariantizations can add algebraic curvature couplings.  On de Sitter such terms collapse to a renormalization of the Maxwell coefficient and a shift of the effective massive pole.  The formulas above correspond to the minimal action \eqref{eq:action}.

\section{Conclusions}

We constructed the transverse retarded BP Green function in four-dimensional de Sitter spacetime by exact Maxwell--massive factorization.  The invariant kernel is controlled by a hypergeometric master function whose common Maxwell and massive $1/y$ singularities cancel.  The one-curl electromagnetic response retains a null-cone delta distribution and a smooth interior tail, while the smooth Maxwell master tail is annihilated by the curl, leaving a physical tail generated entirely by the massive sector.

The local coefficients agree among the hypergeometric, Hadamard, and flat Bessel expansions.  The exact static solution cancels the Coulomb divergence, has the correct flat limit, and obeys a generalized Gauss law.  Its global completion requires antipodal opposite charges and smooth factorized initial data.  These results establish a self-consistent classical causal framework without making claims about quantum Lee--Wick unitarity or de Sitter vacuum selection.

\section*{Acknowledgment}
XYC is supported by National Science Foundation of China (No.~W2533026).  ChatGPT was used for language editing and assistance
with literature searches; the authors checked the calculations, statements,
and bibliographic data and take responsibility for the final manuscript.
%\end{acknowledgments}

\appendix
\section{Normalization audits}
\label{app:audits}

\subsection{Hypergeometric branch cut}
\label{app:branchcut}

Euler's transformation gives
\begin{equation}
 {}_2F_1(a,b;3;u)=(1-u)^{-2}{}_2F_1(\alpha,\beta;3;u),
\end{equation}
where $\alpha=1/2-\nu$, $\beta=1/2+\nu$, and $\alpha+\beta=1$.  Euler's integral representation is
\begin{multline}
 {}_2F_1(\alpha,\beta;3;u)=
 \frac{2}{\Gamma(\beta)\Gamma(3-\beta)}\\
 \times\int_0^1\!\dd t\,
 t^{\beta-1}(1-t)^{2-\beta}(1-ut)^{-\alpha}.
 \label{eq:eulerintegral}
\end{multline}
For $u>1$, only the interval $t>1/u$ crosses the cut.  Since the Feynman prescription is $u-i0$,
\begin{equation}
 (1-(u-i0)t)^{-\alpha}
 =e^{-i\pi\alpha}(ut-1)^{-\alpha},\qquad t>1/u.
\end{equation}
Taking $-2$ times the imaginary part of Eq.~\eqref{eq:eulerintegral}, including
the real factor $(u-1)^{-2}$ from Euler's transformation, yields the main-text
integral Eq.~\eqref{eq:Dnu} and proves the discontinuity relation
Eq.~\eqref{eq:Ddiscontinuity}. Hence, the massive retarded master tail is 
\begin{equation}
 \gamma_{M,\mathrm{tail}}^{\ret}= i\frac{3H^4\Gamma(a)\Gamma(b)}{64\pi^2M^2}
 \theta_+\theta(u-1)\mathcal D_\nu(u).
\end{equation}
The sign is fixed by $u_F=u-i0$; using $u+i0$ would give the opposite
retarded discontinuity.  The logarithmic continuation at $u=1^+$ reproduces
the main-text cone value Eq.~\eqref{eq:Dnucone}:
\begin{equation}
 \mathcal D_\nu(1^+)=
 \frac{2\pi\mu(2+\mu)}{\Gamma(a)\Gamma(b)},
\end{equation}
and hence
\begin{equation}
 \gamma_{M,\mathrm{tail}}^{\ret}(1^+)
 = i \theta_+ \frac{3H^2(2+\mu)}{32\pi}.
\end{equation}

\subsection{Static hypergeometric normalization}
\label{app:staticnormalization}

With $x=H^2r^2$, the massive static equation becomes
\begin{equation}
 x(1-x)\Phi_{M,xx}+\frac32(1-x)\Phi_{M,x}
 -\frac\mu4\Phi_M=0.
 \label{eq:staticxode}
\end{equation}
The parameters $\alpha_s+\beta_s=1/2$ and
$\alpha_s\beta_s=\mu/4$ reduce Eq.~\eqref{eq:staticxode} to the hypergeometric equation.  The solution in Eq.~\eqref{eq:staticmassive} behaves as $1-x$ at the horizon, thus defines a regular one-form in horizon-regular coordinates.

Near $x=0$, the connection formula gives
\begin{multline}
 (1-x){}_2F_1\left(\frac32-\alpha_s,
 \frac32-\beta_s;2;1-x\right)\\
 =C_{\alpha_s\beta_s}x^{-1/2}
 -\frac{2\sqrt\pi}{\Gamma(1/2+\alpha_s)
 \Gamma(1/2+\beta_s)}\\
 + \frac{2\sqrt\pi x^{1/2}}{\Gamma(1/2-\alpha_s)\Gamma(1/2-\beta_s)}+ O(x).
\end{multline}
Because $x^{-1/2}=(Hr)^{-1}$, the prefactor in
Eq.~\eqref{eq:staticmassive} produces exactly $q/(4\pi r)$  and 
\begin{equation}
 \Phi_M=\frac{q}{4\pi r}+\Phi_M^{(0)}
 +\frac{qM^2}{8\pi}r+O(r^2).
\end{equation}
This is the main-text expansion Eq.~\eqref{eq:staticnearorigin}.
The constant connection coefficient gives
\begin{align}
 \Phi_P(0)&=-\frac{q H}{4\pi} -\Phi_M^{(0)}\nonumber \\
 &=\frac{qH}{4\pi}\left[
 2\frac{\Gamma(3/2-\alpha_s)\Gamma(3/2-\beta_s)}
 {\Gamma(1/2+\alpha_s)\Gamma(1/2+\beta_s)}-1\right],
\end{align}
confirming Eq.~\eqref{eq:staticcentral}. 

In the flat limit $H\to 0$, the imaginary parts of $\alpha_s, \beta_s$ go to infinity
\begin{equation}
\alpha_s, \beta_s = \frac{1\pm i\sqrt{4\mu -1}}{4}, \quad \mu=\frac{M^2}{H^2}\to \infty.
\end{equation}
The asymptotic expansion (see Eq.(5.11.9) of ~\cite{NIST:DLMF}) with fixed real part is 
\begin{equation}
|\Gamma (x + i y)| \sim \sqrt{2\pi}|y|^{x - (1/2)} e^{-\pi|y|/2}, \quad, y\to\infty.
\end{equation}
Therefore, we obtain
\begin{equation}
\Phi_P(0)\overset{H\to 0}{\longrightarrow} \frac{q M}{4\pi}.
\end{equation}

\end{document}